\documentclass[prd,twocolumn,nofootinbib,showpacs,superscriptaddress]{revtex4-1}

\usepackage{amsfonts}
\usepackage{amsmath}
\usepackage{amssymb}
\usepackage{bm}
\usepackage{dcolumn}
\usepackage{graphicx}
\usepackage{graphics}
\usepackage[latin1]{inputenc}
\usepackage{latexsym}
\usepackage{rotating}
\usepackage{hyperref}
\usepackage[all]{hypcap}
\usepackage{xspace} 
\usepackage[usenames]{color}

\usepackage{ulem}
\definecolor {darkgreen}{rgb}{0.2,0.7,0.2}

\newcommand{\be}{\begin{equation}}
\newcommand{\ba}{\begin{eqnarray}}
\newcommand{\ee}{\end{equation}}
\newcommand{\ea}{\end{eqnarray}}

\DeclareMathAlphabet{\scr}{OT1}{rsfs}{m}{n} 
\newcommand\bw{\begin{widetext}}
\newcommand\ew{\end{widetext}}

\newcommand{\nn}{\nonumber}

\newcommand{\F}{{\cal F}} 
\newcommand{\K}{{\cal K}} 
\newcommand{\Edot}{\dot{\cal E}} 
\newcommand{\Bdot}{\dot{\cal B}}

\newcommand{\chid}{\chi^{\scriptstyle \sf d}} 
\newcommand{\Eq}{{\cal E}^{\scriptstyle \sf q}} 
\newcommand{\Bq}{{\cal B}^{\scriptstyle \sf q}} 
\newcommand{\Fd}{{\cal F}^{\scriptstyle \sf d}} 
\newcommand{\Fo}{{\cal F}^{\scriptstyle \sf o}} 
\newcommand{\Kd}{{\cal K}^{\scriptstyle \sf d}} 
\newcommand{\Ko}{{\cal K}^{\scriptstyle \sf o}} 
\newcommand{\Edotq}{\dot{\cal E}^{\scriptstyle \sf q}} 
\newcommand{\Bdotq}{\dot{\cal B}^{\scriptstyle \sf q}} 
\newcommand{\Fdotd}{\dot{\cal F}^{\scriptstyle \sf d}} 
\newcommand{\Fdoto}{\dot{\cal F}^{\scriptstyle \sf o}} 
\newcommand{\Kdotd}{\dot{\cal K}^{\scriptstyle \sf d}} 
\newcommand{\Kdoto}{\dot{\cal K}^{\scriptstyle \sf o}} 
\newcommand{\m}{{\sf m}} 
\newcommand{\eq}{e^{\scriptstyle \sf q}} 
\newcommand{\bq}{b^{\scriptstyle \sf q}} 
\newcommand{\ehatq}{\hat{e}^{\scriptstyle \sf q}}

\newcommand{\bhatq}{\hat{b}^{\scriptstyle \sf q}} 
\newcommand{\kd}{k^{\scriptstyle \sf d}} 
 
\newcommand{\ko}{k^{\scriptstyle \sf o}} 
 
\newcommand{\fd}{f^{\scriptstyle \sf d}} 
\newcommand{\fo}{f^{\scriptstyle \sf o}} 
\newcommand{\cc}[1]{c^{\scriptstyle \sf #1}} 
\newcommand{\gam}[1]{\gamma^{\scriptstyle \sf #1}} 
\newcommand{\ccdot}[1]{\dot{c}^{\scriptstyle \sf #1}} 
\newcommand{\gamdot}[1]{\dot{\gamma}^{\scriptstyle \sf #1}}

\newcommand{\ext}{{\mbox{\tiny ext}}}
\newcommand{\orb}{{\mbox{\tiny orb}}}

\newcommand{\E}{{\cal E}} 
\newcommand{\B}{{\cal B}} 

\newcounter{subsubsubsection}[subsubsection]

\begin{document}
\title{Improved next-to-leading order tidal heating and torquing of a Kerr black hole}

\author{Katerina Chatziioannou}
\affiliation{eXtreme Gravity Institute, Department of Physics, Montana State University, Bozeman, Montana 59717, USA}

\author{Eric Poisson}
\affiliation{Department of Physics, University of Guelph, Guelph, Ontario, NIG 2W1, Canada}

\author{Nicol\'as Yunes}
\affiliation{eXtreme Gravity Institute, Department of Physics, Montana State University, Bozeman, Montana 59717, USA}

\date{\today}


%
\begin{abstract}

We calculate the energy and angular-momentum fluxes across the event horizon of a tidally deformed, rapidly rotating black hole to next-to-leading order in the curvature of the external spacetime. These are expressed in terms of tidal quadrupole moments and their time derivatives, which provide a characterization of a generic tidal environment. As an application of our results, we provide an expression for the energy and angular-momentum fluxes across the horizon when the black hole is a member of a binary system on a slowly moving, quasicircular orbit. Our expressions are accurate to 1.5 post-Newtonian order beyond the leading-order fluxes, but they are valid for arbitrary mass ratios. We compare our results to those previously obtained in the case of an extreme mass ratio binary, and find that they do not agree at the 1.5 post-Newtonian order. We investigate a number of possible sources for this discrepancy, but are ultimately unable to resolve it.

\end{abstract}

\maketitle

\section{Introduction}

Astrophysically realistic black holes (BHs) are never in isolation. From the viewpoint of a given \emph{background} BH, the external universe induces gravitational perturbations that \emph{heat and torque} the background BH. This heating and torquing is a flux of energy and angular momentum across the background BH's horizon that lead to a change in its mass and spin. These fluxes are sometimes called \emph{horizon fluxes}~\cite{Yunes:2009ef,PhysRevD.83.044044} or \emph{BH absorption}~\cite{Mino:1997bw} to distinguish them from the fluxes associated with gravitational radiation carried out to infinity.

The horizon fluxes can be computed analytically by integrating the Teukolsky equation for the Newman-Penrose (NP) scalar $\psi_0$, assuming that the effect of the external universe is small~\cite{Poisson:2004cw}. In these circumstances, $\psi_0$ can be expanded in powers of the ratio of the background BH's mass to the radius of curvature of the external universe, which can be parametrized with electric and magnetic tidal tensors. The horizon fluxes can then be computed by evaluating $\psi_0$ at the horizon and performing some operations on it. 

Until recently, the calculation of the horizon fluxes for generic, slowly varying tidal environments had only been carried out to leading order in an expansion in inverse powers of the radius of curvature of the external universe~\cite{Poisson:2004cw,Yunes:2005ve,Comeau:2009bz}. In~\cite{Chatziioannou:2012gq} we calculated these fluxes to next-to-leading order. Here, we improve on these results in two ways: (i) we correct the calculation of the horizon fluxes for binary BHs in a slowly moving, quasicircular orbit at 1.5 post-Newtonian (PN) order, completing our previous computation in~\cite{Chatziioannou:2012gq}, and (ii) we provide ready-to-use flux formulas for comparisons with numerical relativity and for direct use in gravitational-wave modeling.

Ready-to-use expressions are useful because they enable the construction of accurate templates for the gravitational waves emitted by inspiraling BH binaries. This calculation requires knowledge of how the orbit decays due to the loss of energy and angular momentum to the waves. Through a balance law~\cite{lrr-2014-2}, the rates of change of the orbital binding energy and the angular momentum are related to the energy and angular-momentum fluxes out to infinity and into the BH's horizons. For a quasicircular binary composed of spinning BHs, the leading-order term in a PN expansion\footnote{The PN approximation is one in which the field equations are solved as an expansion in small velocities (relative to the speed of light) and weak fields. A term of relative ${\cal{O}}(V^{2A})$ is said to be of $A$th PN order.} of the energy horizon flux is proportional to $V^{15}$, where $V$ is the orbital velocity. This corresponds to a $2.5$ PN order correction relative to the leading-order (quadrupole) energy flux radiated out to infinity, which is proportional to $V^{10}$. In the test-particle limit and for quasicircular orbits, the horizon energy flux into a spinning BH is known to $20$PN order relative to the $V^{15}$ leading-order horizon flux~\cite{Shibata:1994jx,Tanaka:1997dj,Tagoshi:1997jy,Fujita:2014eta,Shah:2014tka}. Here we provide expressions for the horizon energy and angular-momentum fluxes accurate through $1.5$PN order relative to the leading-order horizon flux, i.e.~up to $V^{18}$, but valid for an \emph{arbitrary} mass ratio. These expressions would aid in the construction of waveform templates for comparable-mass, spinning BH quasicircular inspirals. 

A surprising outcome of our calculations is that we do not find agreement between our results in the limit of an extreme mass ratio, and the test-particle calculation of~\cite{Tagoshi:1997jy} at 1.5PN order. We do find agreement at leading-order and at 1PN order. We describe a number of possible culprits for this discrepancy, but ultimately we fail to resolve it. We must thus, unfortunately, leave this question open for the time being. 

\section{Formalism}
\label{sec:horfluxes} 

Consider a perturbed Kerr BH with mass $M$, spin angular momentum $J = M a$, and dimensionless spin parameter $\chi = J/M^{2}$.  The horizon fluxes of energy and angular momentum can be computed from (see~\cite{Chatziioannou:2012gq} for more details)
\begin{align}\label{Mdot-def}
\left<\frac{d M}{dv}\right> &= \frac{r^2_+ + a^2}{4 \kappa} \sum _{m } \left[ 2 \kappa \int \langle |\Phi^m_+|^2 \rangle \sin{\theta} d\theta \right. \nn \\
&- \left. i m \Omega_H \int \langle \bar{\Phi}^m_+ \Phi^m_- - \Phi^m_+ \bar{\Phi}^m_- \rangle \sin{\theta} d\theta \right] \,,
\end{align}
and
\begin{align}\label{Jdot-def}
\left<\frac{d J}{dv}\right> = - \frac{r^2_+ + a^2}{4 \kappa} \sum _{m \neq 0} (i m) \int \langle \bar{\Phi}^m_+ \Phi^m_- - \Phi^m_+ \bar{\Phi}^m_- \rangle \sin{\theta} d\theta \,,
\end{align}
where $(v,r,\theta,\psi)$ are ingoing Kerr coordinates, while $\kappa = (r_+ - M)/(r^2_+ + a^2)$ and $\Omega_H = a/(r^2_+ + a^2)$ are the surface gravity and the angular velocity of the unperturbed BH, respectively~\cite{Poisson:2004cw}. The \emph{integrated curvatures} $\Phi^m_{\pm}$ (and their complex conjugates $\bar{\Phi}^m_{\pm}$) are defined through
\ba
\label{Phimplus}
\Phi^m_+(v,\theta) &=& e^{\kappa v} \int_{v}^{\infty} e^{-(\kappa - i m \Omega_H) v'} \Psi^m(v',\theta) dv'\label{phi+} \,, 
\\
\label{Phimminus}
\Phi^m_-(v,\theta) &=&  \int^{v}_{-\infty} e^{ i m \Omega_H v'} \Psi^m(v',\theta) dv' \label{phi-}\,,
\ea
where $\Psi$ is the \emph{Teukolsky potential}, defined by
\be\label{Psi-def}
\Psi(v,\theta,\psi) 
= -\frac{\Delta^2}{4(r^2+a^2)^2}\psi_0 \bigg|_{r=r_{+}}\,,
\ee
where $\Delta=r^2-2Mr+a^2$, $r_{\pm}=M\pm \sqrt{M^2-a^2}$, and $\psi_0$ is one of the NP scalars in the Kinnersley tetrad. The axial symmetry of the Kerr solution allows us to decompose $\Psi$ in decoupled azimuthal modes, 
\begin{align}
\Psi(v, \theta, \psi) = \sum_m \Psi ^m (v, \theta) e^{i m \psi} .
\end{align}
The first law of BH mechanics allows us to also compute the rate of change of the horizon area via
\be\label{Adot-def}
\frac{\kappa}{8 \pi} \left<\frac{d A}{dv}\right> = \frac{r^2_+ + a^2}{2} \sum _{m}  \int \langle |\Phi^m_+|^2 \rangle \sin{\theta} d\theta \,.
\ee

The NP scalar $\psi_{0}$ can be computed as an expansion in inverse powers of the radius of curvature of the external universe. Working in Fourier space, we express it as~\cite{Chatziioannou:2012gq} 
\begin{align}\label{psi0-f-dec}
\tilde{\psi}_0 &=  \sum_{\ell m} \;  \tilde{z}_{\ell m}(\omega) R_{\omega\ell m}(r) \; {}_{2}S^{\omega\ell m}(\theta) e^{i m \psi}\,,
\end{align}
where $R_{\omega\ell m}(r)$ are functions that satisfy the radial Teukolsky equation, ${}_{2}S^{\omega\ell m}(\theta)$ are spin-weight $+2$ spheroidal harmonics, and $\tilde{z}_{\ell m}(\omega)$ are complex amplitudes. A calculation of the horizon fluxes requires the determination of  these ingredients, which are then inserted in the expressions of the integrated curvatures before substitution into the flux formulas.

\section{Asymptotic matching}
\label{sec:matching}

The functions $R_{\omega\ell m}(r)$ must satisfy the radial Teukolsky equation, and they must be regular at the BH's horizon. Because the differential equation is homogeneous, the regular solution is determined up to an overall multiplicative constant which can be chosen arbitrarily. The information about the tidal environment is then encoded in the amplitudes $\tilde{z}_{\ell m}(\omega)$, which must be determined. We adopt the following strategy. 

In Appendix~\ref{sec:slowrot} we construct the metric of a slowly rotating BH that is placed in a generic, time-dependent tidal environment characterized by quadrupole moments $\E_{ab}(v)$ and $\B_{ab}(v)$. In this computation the BH's dimensionless angular momentum $\chi$ is assumed to be small, and all equations are linearized with respect to $\chi$. The calculation generalizes~\cite{Poisson:2014gka} to account for the time-dependence of the tidal moments, whose derivatives with respect to $v$ enter in a 1.5PN calculation of the horizon fluxes. The metric of the perturbed BH is next used to compute the NP scalar $\psi_0$, which is then evaluated in the asymptotic region $r \gg M$. This expression is exploited to fix the normalization of $R_{\omega\ell m}(r)$ and determine the amplitudes $\tilde{z}_{\ell m}(\omega)$ in terms of the tidal moments.   

It may appear objectionable that a $\psi_0$ calculated to linear order in $\chi$---the one obtained in Appendix~\ref{sec:slowrot}---is used to determine the asymptotic behavior (and therefore the amplitude of each mode) of a $\psi_0$ calculated to all orders in $\chi$---the one that appears in the flux formulas of Sec.~\ref{sec:horfluxes}. Does not the asymptotic behavior of $\psi_0$ contain terms of higher order in $\chi$? The answer to this objection, the key to a successful implementation of our strategy, goes as follows. First, our 1.5PN calculation of the fluxes requires amplitudes $\tilde{z}_{\ell m}(\omega)$ that can be determined from the leading-order asymptotic behavior of $R_{\omega\ell m}(r)$ together with subleading terms of relative order $M/r$; additional terms of order $(M/r)^2$ and beyond are not required. Second, a study of the Teukolsky equation [see Appendix~\ref{sec:MST}, especially Eqs.~(\ref{psi0_expanded1}) and (\ref{psi0_expanded2})] reveals that once the leading-order asymptotic term in $R_{\omega\ell m}(r)$ is chosen to be independent of $\chi$, the subleading term of order $M/r$ is {\it necessarily linear in $\chi$}; higher-order terms in $\chi$ appear only in the additional terms of order $(M/r)^2$ and beyond. These observations therefore imply that a 1.5PN calculation of the horizon fluxes requires amplitudes $\tilde{z}_{\ell m}(\omega)$ that can be determined from the asymptotic behavior of $\psi_0$ calculated to first order in $\chi$. This information can be provided by the calculation presented in Appendix~\ref{sec:slowrot}.     

The final outcome of this exercise, in which we match Eq.~\eqref{radial_asymp} to Eq.~\eqref{psi0-f-dec}, is a radial function normalized by 
\be
R_{\omega 2m}(r) \sim 1 + i \omega \left(\frac{r}{3} + 2 M \ln{\frac{r}{2M}} + M - \frac{\pi^2}{3} i M m \chi \right),\label{as-r}
\ee
and an amplitude given by 
\be
\tilde{z}_{2m} \equiv \tilde{z}_{m,0} + 2 i M \omega \tilde{z}_{m,1}, 
\ee
with  
\begin{align}
\tilde{z}_{m,0} &= \tilde{\alpha}_m + i \tilde{\beta}_m, \\
\tilde{z}_{m,1} &= -m \chi \left(\frac{95}{48} \tilde{\beta}_m - \frac{299}{108} i \tilde{\alpha}_m\right),
\end{align}
where the quantities $\alpha_m$ and $\beta_m$ are defined in terms of ${\cal E}_{ab}$ and ${\cal B}_{ab}$ in Appendix~\ref{sec:slowrot}. 

We note that the metric of Appendix~\ref{sec:slowrot} is not complete, because it does not include terms involving the octupole tidal moments. In an expansion of the metric in inverse powers of the radius of curvature of the external universe, the octupole moments do appear at the same order as terms involving the time derivative of the quadrupole tidal moments. Nevertheless, the octupole moments can be ignored, because they appear only in the $\ell=3$ mode of $\psi_0$, which does not contribute to the horizon fluxes at $1.5$PN order.

\section{Teukolsky function evaluated on the horizon}

The radial function can be decomposed as $R_{\omega 2m} = R_{2m,0} + \omega M R_{2m,1} + O(\omega^2)$ with the asymptotic behavior of each term obtained from Eq.~\eqref{as-r}. The components were obtained in \cite{Chatziioannou:2012gq} by solving the radial Teukolsky equation order by order in $\omega$. We have  
\begin{align}
R_{2m,0} &\!=\! A_{2m} x^{-2} (1\! +\! x)^{-2} F(-4,1;  2 i m \gamma-1 ; \!-x)\label{0-sol}, 
\end{align}
where $ F(a,b;c;z)$ is the hypergeometric function. The constant $A_{2m}$ is determined by ensuring that the asymptotic behavior of this solution matches Eq.~\eqref{as-r}:
\begin{align}
A_{2m} &\!=\! - \frac{i}{6} m \gamma \left(1 + i m \gamma \right) \left(1 + 4 m^{2} \gamma^{2}\right)\,,
\label{A2m-const}
\end{align}
where $\gamma := a/(r_+-r_-)$. We also have 
\begin{align}
R_{2m,1} \!=\! A \; R_{2m,0} + R_{2m,p},\label{1st-sol}
\end{align}
where $A$ is a constant and $R_{2m,p}$ is the particular solution given in Eqs.~(81) and (82) of~\cite{Chatziioannou:2012gq}.

As first noted in~\cite{Chatziioannou:2012gq}, to leading order the asymptotic behavior of $R_{2m,1}$ is $ir/3$, thus satisfying Eq.~\eqref{as-r}. The subleading behavior fixes the constant $A$. This information was not yet available at the time of~\cite{Chatziioannou:2012gq}, and we made the choice of fixing $A$ through the requirement that the NP scalar be regular in the $\chi \to 0$ limit. We must, however, determine $A$ by demanding that Eq.~\eqref{1st-sol} agrees with Eq.~\eqref{as-r}:
\begin{align}
A &=  2 i \left[\psi^{(0)}\left(3 + i m \frac{\chi}{\sigma}\right) + \gamma_E+ \ln{\sigma}\right] + \frac{i}{3} (4 + 5 \sigma) \nn\\
&\!-\! 2 \frac{1 + \sigma}{m \chi}\! +\! \frac{m}{3} (4 \!+\! \pi^2) \chi\! -\!2 i \frac{1 + \sigma}{2 \sigma\! +\! i m \chi}\!-\!  4 m \frac{(1 + \sigma) \chi}{\sigma^2 \!+\! m^2 \chi^2}.
\end{align}
where $\gamma_E$ is the Euler gamma, $\psi^{(n)}(x)$ is the polygamma function, and $\sigma=\sqrt{1-\chi^2}$. With this result we can evaluate $\psi_0$ at the horizon and retrace our steps from~\cite{Chatziioannou:2012gq} to calculate the horizon fluxes.

\section{Horizon fluxes} 
\label{hf}

The calculation of the horizon fluxes from $\psi_0$ is described in detail in~\cite{Chatziioannou:2012gq}. Here we omit details and directly present the final results. Defining the invariants
\begin{align}
E_1 = {\cal{E}}_{ab}{\cal{E}}^{ab}  \,,  \quad \qquad &B_1 = {\cal{B}}_{ab}{\cal{B}}^{ab}  \,,\\
E_2 = {\cal{E}}_{ab} s^b {\cal{E}}^{a}_{c} s^c \,, \qquad &B_2 = {\cal{B}}_{ab} s^b {\cal{B}}^{a}_{c} s^c  \,,\\
E_3 = ({\cal{E}}_{ab} s^a s^b)^2  \,, \qquad  &B_3 = ({\cal{B}}_{ab} s^a s^b)^2  \,,\\
E_4 = \epsilon_{pqc}{\cal{E}}^{pa}\dot{{\cal{E}}}^{q}_{a} s^c \,, \qquad &B_4 = \epsilon_{pqc}{\cal{B}}^{pa}\dot{{\cal{B}}}^{q}_{a} s^c \,,\\
E_5 = \epsilon_{pqc}{\cal{E}}^{p}_{a}\dot{{\cal{E}}}^{q}_{b} s^a s^b s^c  \,, \quad &B_5 = \epsilon_{pqc}{\cal{B}}^{p}_{a}\dot{{\cal{B}}}^{q}_{b} s^a s^b s^c \,,
\end{align}
where $s^a=(0,0,1)$ is the direction of the BH spin and 
\begin{align}
A_m &\equiv \frac{1}{2}\left[\psi^{(0)}\left(3+i m \frac{\chi}{\sigma}\right)+\psi^{(0)}\left(3-i m \frac{\chi}{\sigma}\right)\right],\\
B_m &\equiv \frac{1}{2i}\left[\psi^{(0)}\left(3+i m \frac{\chi}{\sigma}\right)-\psi^{(0)}\left(3-i m \frac{\chi}{\sigma}\right)\right],
\end{align}
we find
\begin{align}\label{Mdotffinal}
\langle \dot{M} \rangle &= \langle \dot{M}^{(5)} \rangle\,, 
\end{align}
\be\label{Jdotffinal}
\langle \dot{J} \rangle=\langle \dot{J}^{(4)} \rangle+\langle \dot{J}^{(5)}_1 \rangle+\langle \dot{J}^{(5)}_2 \rangle+\langle \dot{J}^{(5)}_3 \rangle\,,
\ee
\be\label{Adotffinal}
\langle \dot{A} \rangle=-\frac{8\pi\chi}{\sigma}\left[\langle \dot{J}^{(4)} \rangle+\langle \dot{J}^{(5)}_1 \rangle+\langle \dot{J}^{(5)}_3 \rangle\right]+\langle \dot{A}^{(5)}_2 \rangle\,,
\ee
where

\allowdisplaybreaks
\begin{widetext}
\begin{subequations}
\begin{align}
\langle \dot{M}^{(5)} \rangle &=\frac{2 M^5 \chi}{45}\left[ -4\left(3\chi^2 + 1 \right) \langle E_4 + B_4 \rangle + 15 \chi^2\langle E_5+ B_5 \rangle \right]\,,
\\
\langle \dot{J}^{(4)} \rangle &= - \frac{2 M^5 \chi}{45} [8 (1 + 3 \chi^2) \langle E_1 + B_1 \rangle - 3 (4 + 17 \chi^2) \langle E_2 + B_2 \rangle + 15 \chi^2 \langle E_3 + B_3 \rangle ]\,,
\\
\langle \dot{J}^{(5)}_1 \rangle &= \frac{2 M^6\chi }{135}\left\{ 8 [-5 -4\sigma+ 6(2+ \sigma )  \chi^2 + 9 \chi^4 + 6 (A_2 +\gamma_E +\ln{\sigma}) (1 + 3 \chi^2) ] \langle \dot{E}_1 \!+\! \dot{B}_1 \rangle \right.\nn
\\
&\left.+ 3 [20 + 16 \sigma - (50 + 31 \sigma) \chi^2 - 54 \chi^4 - 32 A_2 (1 + 3 \chi^2) + 2A_1 (4 - 3 \chi^2)-6 (\gamma_E +\ln{\sigma}) (4 + 17 \chi^2)] \langle \dot{E}_2\! +\! \dot{B}_2 \rangle\right.\nn\\
&\left.+3 [8 A_2 (1 + 3 \chi^2) - 2A_1 (4 - 3 \chi^2) +30 (\gamma_E +\ln{\sigma}) \chi^2+ \chi^2 (2 + 7 \sigma + 18 \chi^2)] \langle \dot{E}_3 \!+\! \dot{B}_3\rangle\right\}  \nn\\
&+ \frac{4 M^6\chi }{135}\left\{ 16(3 B_2 - \chi \pi^2) (1 + 3 \chi^2)\langle E_4 \!+\! B_4\rangle +3\left[ \chi \pi^2(4+17\chi^2) + 2 B_1 (4-3\chi^2)-16 B_2 (1+3 \chi^2) \right]\langle E_5 \!+\! B_5 \rangle \right\}\,,
\\
\langle \dot{J}^{(5)}_2 \rangle &= \frac{4 M^6 }{135}\left\{4 [3 (1+ \sigma) + (23+ 39 \sigma) \chi^2 - 6(5 - 3 \sigma) \chi^4]\langle E_4 \!+\! B_4\rangle -3 \chi^2 [29 + 45 \sigma - (38 - 30 \sigma) \chi^2] \langle E_5 \!+\! B_5 \rangle \right\}\,,
\\
\langle \dot{J}^{(5)}_3 \rangle &= \frac{598 M^6 \chi^2}{1215}\left[ 16\left(3\chi^2 + 1 \right) \Big \langle E_4 + \frac{855}{1196}B_4  \Big\rangle - 3(4+17\chi^2) \Big\langle E_5+ \frac{855}{1196}B_5  \Big\rangle \right],
\\
\langle \dot{A}^{(5)}_2 \rangle &= \frac{32 M^6 \pi \chi}{135\sigma}\left\{-8 [3(1 + \sigma) + 8(2+3 \sigma) \chi^2 - 3(5-3 \sigma) \chi^4]\langle E_4 \!+\! B_4\rangle +6 \chi^2 [22 + 30 \sigma - (19 - 15 \sigma) \chi^2] \langle E_5 \!+\! B_5 \rangle \right\}\,.
\end{align}
\end{subequations}
\end{widetext}
The superscripts $(4)$ and $(5)$ give the order of each term in an expansion in powers of $1/{\cal{R}}$, with $\cal R$ denoting the radius of curvature of the external universe.

\section{Circular Binary}

One of the most interesting astrophysical applications of our results is the case of a circular binary with an external BH with mass $M_{\ext}$ and dimensionless spin parameter $\chi_{\ext}$ and a background BH. The angular velocity of the tidal fields in the BH frame is~\cite{Poisson:2014gka} 
\begin{align} \label{Omegaslow}
\Omega &= \epsilon \sqrt{\frac{M_T}{b^3}}\left[ 1 - \frac{1}{2}(3 + \eta) V^2 - \frac{1}{2} \bar{\chi} V^3+{\cal{O}}(V^{4}) \right ] \,,
\end{align}
where $\epsilon = + 1$ ($-1$) if the orbital and spin angular momentum of the unperturbed BH are aligned (antialigned), $\eta  = f f_{\ext}$ is the symmetric mass ratio, $M_{T} = M + M_{\ext}$ is the total mass, $f = M/M_{T}$ and $f_{\ext} = M_{\ext}/M_{T}$ are the mass fractions, $b$ is the orbital separation in harmonic coordinates, $V = (M_{T}/b)^{1/2}$, and $\bar{\chi} \equiv f\left(1+f\right)\chi  + 3 \eta \chi_{\ext}$. 
Equation~\eqref{Omegaslow} corrects Eq.~$(120)$ in~\cite{Chatziioannou:2012gq}, which did not include the $V^{3}$ term.

The angular velocity of the tidal fields is \emph{not} equal to the orbital angular velocity. The latter is given in the PN barycentric frame by  
\begin{equation} 
\omega_{\orb} = \sqrt{\frac{M_{T}}{b^3}} \biggl[ 1 - \frac{1}{2}(3-\eta) V^2 
- \frac{1}{2} \tilde{\chi} V^3 + {\cal{O}}(V^4) \biggr], 
\end{equation} 
where $\tilde{\chi} \equiv (2f^{2} + 3 \eta) \chi +  (3 \eta + 2f_\ext^{2}) \chi_\ext$. Even though functionally $\Omega$ looks similar to $\omega_{\orb}$, these expressions are clearly not the same because $\tilde{\chi} \neq \bar{\chi}$.

Evaluation of the horizon fluxes when the background BH is a member of a binary requires expressions for the tidal 
fields that are accurate to the appropriate PN order. The tidal fields were obtained to $1$PN order in~\cite{Taylor:2008xy}, and extended to $1.5$PN order in~\cite{Poisson:2014gka}; they can be used to compute the horizon fluxes to ${\cal{O}}(V^{3})$ relative to the leading-order horizon absorption term.  The relevant electric tidal fields are 
\begin{align}
&\frac{1}{2}({\cal{E}}_{11} \!+\! {\cal{E}}_{22}) = - \frac{M_{\ext}}{2b^3}\!\left[1 + \frac{f}{2}V^2 - 6 f_{\ext}\chi_\ext V^3  + {\cal{O}}(V^4)\right] \!\label{Eslow-beg}, \\
&\frac{1}{2}({\cal{E}}_{11} \!-\!{\cal{E}}_{22}) = - \frac{3M_{\ext}}{2b^3}\!\left[1 + \frac{f-4}{2}V^2 - 2f_{\ext}\chi_\ext V^3
\right. 
\nn \\ 
&\left. \qquad\qquad\qquad+ {\cal{O}}(V^4)\right]\! \cos{2 \Omega t}, \\
&{\cal{E}}_{12}\! =\! - \frac{3M_{\ext}}{2b^3}\!\left[1 \!+\! \frac{f\!-\!4}{2}V^2 \!-\! 2f_{\ext}\chi_\ext V^3 
\!+\! {\cal{O}}(V^4)\right] \!\sin{2 \Omega t} \,, 
\end{align}
and the relevant magnetic tidal fields are
\begin{align}
{\cal{B}}_{13} &=  -\frac{3 M_{\ext}}{b^3} V \left(1-f_{\ext}\chi_\ext V\right)\cos{\Omega t} + {\cal{O}}(V^3)\,,\\
{\cal{B}}_{23} &=  -\frac{3 M_{\ext}}{b^3} V \left(1-f_{\ext}\chi_\ext V\right)\sin{\Omega t} + {\cal{O}}(V^3)\label{Bslow-end}\,,
\end{align}
improving Eqs.~$(122)$--$(126)$ of~\cite{Chatziioannou:2012gq}. Defining
\begin{align}
C_V&= -\frac{16}{5}M^2 f^2 \eta^2 (1+\sigma)V^{12} \biggl\{ 1 + 3\chi^2 \nn
\\
&-\left[3+\frac{51}{4}\chi^2 - (1+3\chi^2) f \right] V^2  \nn 
\\ & +\! \left\{\!\frac{8}{3}\epsilon f (1+3\chi^2)(\pi^2\chi-3B_2) -\frac{3}{2} f_{\ext} \chi_{\ext}(4+7\chi^2)\right.\nn
\\
&\left. -\frac{4}{27}\epsilon f \chi \left[ 362+135\sigma \!+\!(762+81\sigma)\chi^2   \right]  \!  \right\}\! V^3\! +\! {\cal{O}}(V^4)\! \biggr\},\label{CJ-def}
\end{align}
the energy and angular-momentum flux become
\begin{align}
\left< \frac{dJ}{dv} \right> &= (\Omega_H-\Omega) C_V \label{Jdotslow},
\\
\left< \frac{d M}{dv} \right> &= \Omega  (\Omega_H-\Omega) C_V \label{Mdotslow},
\end{align}
respectively, while the change in horizon area is simply
\begin{align}
\left< \frac{d A}{dv} \right> &= -\frac{8\pi}{\kappa}(\Omega_H-\Omega)^2 C_V\label{Adotslow}.
\end{align}
These expressions correct Eqs.~$(127)-(129)$ in~\cite{Chatziioannou:2012gq}, which miscalculated the $V^{15}$, $V^{18}$, and $V^{15}$ terms, respectively. 

Equations~\eqref{Jdotslow}-\eqref{Adotslow} are presented in their factorized form, in that the fluxes are all proportional to $\Omega_H - \Omega$. This form includes more terms than what we are formally allowed to keep. For example, the energy flux in Eq.~\eqref{Mdotslow} contains terms proportional to $V^{19}$ to $V^{24}$, none of which we are formally allowed to retain, since Eq.~\eqref{CJ-def} has uncontrolled remainders of ${\cal{O}}(V^{4})$. However, these factorized expressions make it clear that the fluxes vanish in the case of corotation, which we expect on physical grounds. Comparison with numerical simulations could determine whether the factorized forms are more accurate than the fully expanded forms.  
 
The expressions for the horizon fluxes computed above have been written in terms of the variable $V = \sqrt{M_{T}/b}$, which is clearly coordinate dependent through the harmonic orbital separation $b$.  A more meaningful expression may be obtained if we adopt $x = (M_{T}\omega_{\orb})^{1/3}$ as a coordinate-invariant expansion parameter. The relation is provided by  
\begin{equation} 
V = x \biggl[ 1 + \frac{1}{6} (3- \eta) x^2 
+ \frac{1}{6} \tilde{\chi} x^3 + {\cal{O}}(x^4) \biggr], 
\end{equation} 
while the angular velocity of the tidal field is
\begin{equation} 
\Omega =\epsilon \frac{x^3}{M_T} \biggl[ 1 - \eta x^2 
+ \frac{1}{2} (\tilde{\chi}-\bar{\chi}) x^3 + {\cal{O}}(x^4) \biggr], \label{Omega}
\end{equation} 

Moreover, Eqs.~\eqref{Jdotslow}-\eqref{Adotslow} are perhaps not in an ideal form yet, because
the time derivatives refer to $v$, an advanced-time coordinate on the BH horizon. 
This is related in a simple way to $\bar{t}$, a time coordinate defined in the local
asymptotic rest frame of the BH. The relation between 
$\bar{t}$ and the PN barycentric time $t$ is given by~\cite{Taylor:2008xy} 
\begin{equation} 
t = \biggl[ 1 + \frac{1}{2} (2f+3f_{\ext})f_{\ext} x^2 + {\cal{O}}(x^{4}) \biggr]
\bar{t},  
\end{equation} 
and it was confirmed in~\cite{Poisson:2014gka} that there are no terms  
at order $x^3$. 

We thus arrive at expressions that could be \emph{directly} implemented in gravitational waveform construction
for comparable-mass, spinning BH binaries in quasicircular orbits. 
Translating the $d/dv$ fluxes to $d/dt$ fluxes and expressing them in terms of $x$, 
we obtain 
\begin{align}
\left< \frac{dJ}{dt} \right> &= (\Omega_H-\Omega) C_x \label{Jdott},
\\
\left< \frac{d M}{dt} \right> &= \Omega  (\Omega_H-\Omega) C_x \label{Mdott},
\\
\left< \frac{d A}{dt} \right> &= -\frac{8\pi}{\kappa}(\Omega_H-\Omega)^2 C_x\label{Adott}.
\end{align}
where now $\Omega$ is given by Eq.~\eqref{Omega} and
\begin{align} 
C_x & = -\frac{16}{5}M^2 f^2 \eta^2 (1+\sigma)x^{12} \biggl\{ 1 + 3\chi^2 \nn
\\
&+\frac{1}{4}\left[3(2+\chi^2)+2f (1+3\chi^2) (2 +3  f) \right] x^2  \nn 
\\ & +\! \left\{\!\frac{8}{3}\epsilon f (1+3\chi^2)(\pi^2\chi-3B_2) -2f\chi(1+3\chi^2)(f-3)\right.\nn
\\
&\left. -\frac{4}{27}\epsilon f \chi \left[ 362+135\sigma \!+\!(762+81\sigma)\chi^2   \right]\right.\nn
\\
&\left.-\frac{1}{2} f_{\ext} \chi_{\ext}(4f_{\ext}-3(1+4f)\chi^2)  \!  \right\}\! x^3\! +\! {\cal{O}}(x^4)\! \biggr\}\label{Cx-def}.
\end{align} 

Eventual comparisons with numerical results on the tidal heating and torquing of a spinning BH will have to clarify the relation between the time coordinate used in the numerical simulation and the PN barycentric time. It may be wiser to adopt a coordinate-invariant parametrization based on the orbital angular velocity $\omega_{\orb}$, which monotonically increases with time because of radiation reaction. An expression for $d\omega_{\orb}/dt$ that includes $1.5$PN terms can be found in Eq.~$(4.14)$ of~\cite{Kidder:1995zr}. Using $x = (M_T\omega_{\orb})^{1/3}$ instead of $\omega_{\orb}$ we find
\begin{align}
\left< \frac{dJ}{dx} \right> &= (\Omega_H-\Omega) C'_x \label{dJdx},
\\
\left< \frac{d M}{dx} \right> &= \Omega  (\Omega_H-\Omega) C'_x \label{dMdx},
\\
\left< \frac{d A}{dx} \right> &= -\frac{8\pi}{\kappa}(\Omega_H-\Omega)^2 C'_x\label{dAdx}.
\end{align}
where
\begin{align}
C'_x & = -\frac{1}{2}M^3 f \eta (1+\sigma)x^{3} \biggl\{ 1 + 3\chi^2 \nn
\\
&+\left[\frac{1}{336}(1247+2481\chi^2)+\frac{5}{4}(3-f)f (1+3\chi^2) \right] x^2  \nn 
\\ & +\! \left\{\!\frac{8}{3}\epsilon f (1+3\chi^2)(\pi^2\chi\!-\!3B_2) \!+\! \frac{7}{12}f\chi(1+3\chi^2)(21\!+\!2f)\right.\nn
\\
&\left. -\frac{1}{12} f_{\ext} \chi_{\ext}\left[-89+14f-21(17-2f)\chi^2\right]\right.\nn
\\
&\left.-\frac{4}{27}\epsilon f \chi \left[ 362+135\sigma \!+\!(762+81\sigma)\chi^2   \right]  \!  \right\}\! x^3\! +\! {\cal{O}}(x^4)\! \biggr\}\label{Cxp-def}.
\end{align}
We recall that the fluxes are here presented in a factorized-resummed form and include uncontrolled PN order terms.  

\section{Small mass ratios}
\label{emri}

The expressions for the horizon fluxes derived here are limited to 1.5PN order, but they are valid for arbitrary mass ratios. On the other hand, Ref.~\cite{Tagoshi:1997jy} uses the formalism of~\cite{Teukolsky:1973ap,Teukolsky:1974yv} to calculate the energy flux to higher PN order (4PN), but the expression is restricted to test particles. Appendix D in~\cite{Tagoshi:1997jy} gives the energy flux across the horizon as a function of $x = (M_T\omega_{\orb})^{1/3}$ when a test particle orbits a Kerr BH. This result truncated to 1.5PN order should be identical to our Eq.~\eqref{Mdott} in the limit of small mass ratios. We find that this is not the case.  

The difference between our Eq.~\eqref{Mdott} and the test-mass result of~\cite{Tagoshi:1997jy} arises at the 1.5PN order, and is given by 
\be
\frac{8}{135} x^{18} \eta^2 \chi^2 [872  +2751 \chi^2 - 72 \pi^2 (1 + 3 \chi^2)]\label{error} .
\ee
Despite the extensive investigations of our calculation described in the following section, we are unable to locate the source of the disagreement.

\section{Discussion and Conclusions}

The discrepancy between our results in the test-mass limit and the results of~\cite{Tagoshi:1997jy} merits further investigation. Below we revisit the individual elements of our calculation and describe how we have checked their validity. 

\subsection{Solution to the Teukolsky equation}

The first ingredient of our calculation---and indeed of the calculation of~\cite{Tagoshi:1997jy}---is a homogeneous solution to the Teukolsky equation to the appropriate order in $M\omega\sim M/{\cal{R}}$. Mano, Suzuki, and Takasugi~\cite{Mano:1996vt} found an exact solution to the homogeneous Teukolsky equation as a series in hypergeometric and Coulomb functions. This solution is utilized in the calculation of~\cite{Tagoshi:1997jy} but not here (or in~\cite{Chatziioannou:2012gq}), because we opted to integrate the Teukolsky equation order by order in $\omega$. 

To test whether our solution to the Teukolsky equation contains errors that could account for the energy flux discrepancy we first substituted it back to the Teukolsky equation and determined that it is indeed a solution. We also repeated our calculations using the series solution of~\cite{Mano:1996vt}. The details are provided in Appendix~\ref{sec:MST}. We find that the flux calculated in this way is identical to Eq.~\eqref{Mdott}, showing that our solution to the Teukolsky equation is correct.

\subsection{Asymptotic matching}
\label{matchingcheck}

With a solution to the homogeneous Teukolsky equation in hand, our next step is to determine its amplitude by examining its asymptotic behavior at infinity. This is obtained in Appendix~\ref{sec:slowrot}, where we construct the perturbed metric of a slowly rotating BH and extract $\psi_0$ from this construction. We have investigated a number of subtleties of the calculation (listed below) that might have led to an incorrect NP scalar at infinity, but without encountering an error. 
\begin{enumerate}
\item We use the metric of a tidally deformed, slowly rotating BH to calculate the asymptotic expression of the NP scalar to all orders in $\chi$. We have shown that corrections in the NP scalar that enter at relative order $M/r$ must be linear in $\chi$, so they are fully captured with a metric linearized in $\chi$. This conclusion is supported by Eq.~(\ref{psi0_expanded2}), which reveals that indeed, all $M/r$ terms are linear in $\chi$. Higher orders in $\chi$ will appear through terms that go as $a^2/r^2\sim \chi^2 M^2/r^2$. Such terms would be necessary in a calculation of the fluxes to next-to-next-to-leading order, but they are not needed here. 
\item We ignore the octupole tidal moments that enter the perturbed metric at the same order as the derivatives of the quadrupole moments. However, octupole moments affect only the $\ell=3$ mode of the NP scalar~\cite{Poisson:2009qj}, which does not contribute to the 1.5PN fluxes~\cite{Chatziioannou:2012gq}. So even though our perturbed metric is not complete at next-to-leading order, it is sufficient for our purposes. 

That the $\ell=3$ mode does not affect the fluxes to next-to-leading order is not obvious; after all it is the next-order mode after the leading $\ell=2$ one. However, as explained in more detail in~\cite{Chatziioannou:2012gq}, the NP scalar needs to be squared and angle averaged over in order to calculate the horizon fluxes. Squaring makes terms obtained by a product of $\ell=3$ modes too high of an order for our purposes, while angle averaging kills any cross terms mixing $\ell=3$ and $\ell=2$ modes. As a consequence, all $\ell=3$ modes drop out of the next-to-leading horizon fluxes.
\item The metric of Eq.~\eqref{metric} is written in light-cone coordinates in which the azimuthal angle $\phi$ is constant on incoming null geodesics. On the other hand, the NP scalar is decomposed in spherical harmonics with an angle $\psi$ that is constant on the ingoing principal congruence of the Kerr spacetime. The mapping between the two angles is given in Eq.~\eqref{phi_vs_psi} to leading order in $\chi$. As we have argued, this relation, which neglects terms of order $\chi^2$ and beyond, is adequate for the computation of the asymptotic behavior of $\psi_0$. 
\item A number of other possible coordinate mismatches have also been explored. For example, the radial coordinate $r_g$ that enters the metric of Eq.~(\ref{metric}) could be related to the $r_{\rm T}$ of the Teukolsky equation by an equation of the form $r_g = r_{\rm T} + k a^2/r_{\rm T} + \cdots$, where $k$ is an unknown constant, and the remaining terms are higher order in $a$. But Eqs.~(\ref{radial_asymp}) show that such a mismatch would have no impact on our results: transforming the expressions from $r_g$ to $r_{\rm T}$ would keep them unchanged, with the mismatch merely contributing to the neglected terms of order $M/r_{\rm T}$. As another example, the advanced-time coordinate $v_g$ of the perturbed metric could differ from the $v_{\rm T}$ of the Teukolsky equation by a term of the form $k' a^2/r + \cdots$. A careful inspection of the developments in Appendix \ref{sec:slowrot} reveals that again, such a mismatch has no impact on our result. 
\item Apart from coordinate differences, matching calculations can suffer from differences in how the spacetime parameters ($M$ and $a$) are defined in each part of the calculation. However, such a difference would appear at leading order in the fluxes. The fact that we only find a discrepancy at relative order $x^3$ indicates that there is a problem with a certain PN expansion, rather than a parameter mismatch.
\end{enumerate}
After this examination we find no reason to suspect the matching procedure and must conclude that it is robust. This conviction is reinforced by the fact that we have verified that the NP scalar of Eq.~\eqref{psi0_1} satisfies the Teukolsky equation to leading order in $\chi$.


\subsection{Tidal fields}

The quadrupole tidal fields caused by a companion BH in a circular binary with the background BH were calculated in~\cite{Poisson:2014gka}. A slowly rotating BH metric that included only quadrupole tidal moments was expanded to 1.5PN order and matched to a PN metric valid to the same order, after both metrics were expressed in the same coordinate system. The result of the matching procedure were the quadrupole tidal fields ${\cal{E}}_{ab}$ and ${\cal{B}}_{ab}$ as a function of the parameters that appear in the metric. 

Two ingredients are missing from the perturbed metric of~\cite{Poisson:2014gka} in order for it to be complete at 1.5PN order: time derivatives of the quadrupole moments $\dot{{\cal{E}}}_{ab}$ and $\dot{{\cal{B}}}_{ab}$, and octupole moments ${\cal{E}}_{abc}$ and ${\cal{B}}_{abc}$. The latter can be safely ignored since the 1.5PN horizon fluxes depend only on the $\ell=2$ mode of the NP scalar; octupole moments and the resulting $\ell=3$ modes enter at higher orders. Moreover, it was argued in~\cite{Poisson:2014gka} that terms proportional to $\dot{{\cal{E}}}_{ab}$ and $\dot{{\cal{B}}}_{ab}$ result only in a phase shift of the tidal fields. As such, they do not affect our flux calculations. We should also note that terms proportional to $\chi\dot{{\cal{E}}}_{ab}$ and $\chi\dot{{\cal{B}}}_{ab}$ were not explicitly included in the analysis of~\cite{Poisson:2014gka}; however, Eq.~\eqref{metric} implies that they make no contribution at 1.5PN order. 

Finally, we should note that the 1.5PN contributions to the tidal fields calculated in~\cite{Poisson:2014gka} depends only on the external BH and not on the background BH, as does the 1PN term. This seemingly curious result can be easily explained: the tidal fields are caused by the external BH and depend on the background BH only through nonlinear interactions between the two BHs. Therefore they have no contribution at 1.5PN order.

We conclude that the tidal fields obtained in~\cite{Poisson:2014gka} are accurate enough for our purpose of obtaining next-to-leading order horizon fluxes and we find no reason to suspect their derivation.

\subsection{Conclusions}

We have calculated the energy and angular-momentum horizon fluxes, as well as the change in horizon area, for a Kerr BH in a circular binary with another BH to next-to-leading order in the curvature of the external spacetime. When taking the test-particle limit of our results we do not recover the results of~\cite{Tagoshi:1997jy}. We have performed a systematic analysis of our calculations in an attempt to locate the cause of the discrepancy, though without success. Apart from the conceptual issues we extensively explored in the previous subsections, we can confidently rule out computational errors: our calculation was performed three times independently, always yielding the same result.

Even though we cannot confidently locate the origin of the discrepancy, a simple observation provides a clue: our result contains factors of $\pi^2$ while the test-particle one does not. These factors originate from the asymptotic behavior of the Teukolsky function calculated in Appendix~\ref{sec:slowrot}; see Eq.~\eqref{radial_asymp} and the subsequent discussion. It then would be reasonable to speculate that the discrepancy originates from the matching procedure, however, we find no further indication that this might be the case.

As a concluding remark, we mention that the results of~\cite{Tagoshi:1997jy} were checked against the numerical results in~\cite{Hughes:1999bq,Hughes:2001jr,Yunes:2008tw,Zhang:2011vha,Yunes:2009ef,Yunes:2010zj} both employing the Mano-Suzuki-Takasugi (MST) machinery of \cite{Mano:1996vt} and with the independent formulation of~\cite{Sasaki:1981kj,Sasaki:1981sx}. Moreover, the analytic calculation of~\cite{Tagoshi:1997jy} was independently verified in~\cite{lrr-2003-6}, further reinforcing confidence in the results of~\cite{Tagoshi:1997jy}. We must unfortunately leave this matter unresolved for the time being.  

\acknowledgments

We thank Ryuichi Fujita, Hiroyuki Nakano, Norichika Sago, Misao Sasaki, Mark Scheel, Hideyuki Tagoshi, Takahiro Tanaka, and Niels Warburton for helpful discussions while working on this problem.  K.C. acknowledges support from the Onassis Foundation. N.Y. acknowledges support from NSF CAREER Grant No. PHY-1250636. E.P. acknowledges support from the Natural Sciences and Engineering Research Council of Canada. 

\appendix

\section{Slowly rotating BH in a time-dependent tidal environment}
\label{sec:slowrot}

In order to specify the asymptotic behavior of the NP scalar when $r \gg M$, we construct the metric of a slowly rotating BH with mass $M$ and dimensionless spin vector $\chi^a$ placed in a tidal environment characterized by quadrupole tidal moments $\E_{ab}(v)$ and $\B_{ab}(v)$. The metric of the deformed BH is calculated in a region that excludes the external matter responsible for the tidal field. We generalize the results of~\cite{Poisson:2014gka} by accounting for the time dependence of the tidal moments; terms proportional to $\dot{\E}_{ab} = d\E_{ab}/dv$ and $\dot{\B}_{ab} = d\B_{ab}/dv$ are now included in the metric, but second-derivative terms are neglected. With this metric in hand, we calculate the NP scalar $\psi_0$ and extract its asymptotic behavior.

\subsection{Tidal potentials} 
\label{sec:potentials} 

The construction of tidal potentials is presented in detail in~\cite{Poisson:2014gka}. Here we summarize the main results, and introduce
new potentials associated with $\dot{\E}_{ab}$ and
$\dot{\B}_{ab}$.  

The potentials are obtained by combining $\chi_a$, $\E_{ab}$,
$\B_{ab}$, and $\Omega^a = [\sin\theta\cos\phi, \sin\theta\sin\phi,
\cos\theta]$ in various irreducible ways, with each potential having
a specific multipole order $\ell$ and a specific parity label (even or
odd).  

The coupling of $\chi_a$ and $\E_{ab}$ produces the pseudotensors
\begin{equation} 
\F_a = \E_{ab} \chi^b, \qquad 
\F_{abc} = \E_{\langle a b} \chi_{c\rangle}, 
\label{F_def} 
\end{equation}  
with angular brackets denoting
symmetrization and trace removal. The coupling of $\chi_a$ and $\B_{ab}$ produces the tensors  
\begin{equation} 
\K_a = \B_{ab} \chi^b, \qquad 
\K_{abc} = \B_{\langle a b} \chi_{c\rangle}.  
\label{K_def} 
\end{equation} 
The independent components of $\E_{ab}$, $\B_{ab}$,
$\F_a$, $\F_{abc}$, $\K_a$, $\K_{abc}$, and $\chi_a$ can be packaged in 
spherical-harmonic coefficients $\Eq_\m$, $\Bq_\m$, $\Fd_\m$,
$\Fo_\m$, $\Kd_\m$, $\Ko_\m$, and $\chid_\m$, respectively. The definitions are
given in Table II of~\cite{Poisson:2014gka}.

The tidal potentials are decomposed in scalar, vector, and tensor
spherical-harmonic functions of the angular coordinates 
$\theta^A = (\theta,\phi)$. The decomposition involves the scalar
harmonics of Table I of~\cite{Poisson:2014gka}, and the even- and odd-parity harmonics of Eqs.~(2.12) and (2.13) of~\cite{Poisson:2014gka}.

The decomposition of the tidal potentials in spherical harmonics is 
described by Eq.~(2.15) of~\cite{Poisson:2014gka}. Together with these we introduce ``dotted potentials'' that are
constructed in an analogous way from $\Edot_{ab} =
d\E_{ab}/dv$ and $\Bdot_{ab} = d\B_{ab}/dv$. For example, 
\begin{align} 
\Edotq &=  \sum_\m \Edotq_\m Y^{2\m}, \quad 
\Fdotd_A = \sum_\m \Fdotd_\m X^{1\m}_A, \nn
\\
\Kdoto_{AB}& = \frac{1}{3} \sum_\m \Kdoto_\m Y^{3\m}_{AB},\nn 
\end{align} 
are dotted potentials, with $\Edotq_\m$, $\Fdotd_\m$, and $\Kdoto_\m$
constructed from $\Edot_{ab}$ and $\Bdot_{ab}$ (and $\chi^a$) in the
manner described in Table II of~\cite{Poisson:2014gka}.  

\subsection{Metric of the deformed BH} 
\label{sec:metric} 

The metric of an isolated, slowly rotating BH of mass $M$ and 
dimensionless spin $\chi$ can be expressed as 
\begin{equation} 
ds^2 = -f_s\, dv^2 + 2\, dvdr + r^2 d\Omega^2 
- 2\frac{2\chi M^2}{r} \sin^2\theta\, dv d\phi, 
\label{background_metric} 
\end{equation} 
where $f_s = 1-2M/r$ and 
$d\Omega^2 = \Omega_{AB} d\theta^A d\theta^B 
= d\theta^2 + \sin^2\theta\, d\phi^2$. The metric is displayed in 
coordinates $(v,r,\theta,\phi)$ that are well behaved on the event
horizon. They are tied to the behavior of incoming null geodesics
that are tangent to converging null cones: 
each surface $v = \mbox{constant}$ is a null hypersurface,
the null generators move with constant values of $\theta$ and $\phi$, and
$-r$ is an affine parameter on each null geodesic~\cite{Poisson:2003wz}. The azimuthal
coordinate $\phi$ differs from $\psi$, which is constant on the ingoing
principal congruence of the Kerr spacetime; the relation is 
\begin{equation} 
\psi = \phi - \chi \frac{M}{r}+{\cal{O}}(\chi^2). 
\label{phi_vs_psi} 
\end{equation} 

The metric of a slowly rotating BH immersed in a tidal field
produced by remote matter is obtained by perturbing Eq.~(\ref{background_metric}). The methods to construct  
the perturbation are described in detail in~\cite{Poisson:2014gka}, in the
case when the time dependence of the tidal moments can be 
neglected.

We continue to work in light-cone coordinates, so that the coordinates
$(v,r,\theta,\phi)$ keep their geometrical meaning in the perturbed
spacetime. This implies that $g_{vr} = 1$, $g_{rr} = 0 = g_{rA}$, so that 
$g_{vv}$, $g_{vr}$, $g_{vA}$, and $g_{AB}$ are the only nonvanishing
components of the metric~\cite{Poisson:2009qj}.  

The perturbed metric is written as 
\bw
\begin{subequations} 
\label{metric} 
\begin{align} 
g_{vv} &= -f_s 
- r^2 \eq_1\, \Eq 
+ \frac{1}{3} r^3 \eq_2\, \Edotq 
- r^2 \ehatq_1\, \chi \partial_\phi \Eq 
+ r^3 \ehatq_2\, \chi \partial_\phi \Edotq 
+ r^2 \kd_1\, \Kd 
+ r^3 \kd_2\, \Kdotd 
- r^2 \ko_1\, \Ko
+ r^3 \ko_2\, \Kdoto, \\ 
g_{vr} &= 1, \\ 
g_{vA} &= \frac{2M^2}{r} \chid_A 
- \frac{2}{3} r^3 \bigl( \eq_4\, \Eq_A - \bq_4\, \Bq_A \bigr) 
+ \frac{1}{3} r^4 \bigl( \eq_5\, \Edotq_A - \bq_5\, \Bdotq_A \bigr) 
- r^3\, \chi \partial_\phi \bigl( \ehatq_4 \Eq_A 
  - \bhatq_4\, \Bq_A \bigr) 
+ r^4\, \chi \partial_\phi \bigl( \ehatq_5 \Edotq_A 
  + \bhatq_5\, \Bdotq_A \bigr) 
\nonumber \\ & \quad \mbox{} 
- r^3 \bigl( \fd_4\, \Fd_A - \kd_4\, \Kd_A \bigr) 
+ r^4 \bigl( \fd_5\, \Fdotd_A + \kd_5\, \Kdotd_A \bigr) 
+ r^3 \bigl( \fo_4\, \Fo_A + \ko_4\, \Ko_A \bigr)
+ r^4 \bigl( \fo_5\, \Fdoto_A + \ko_5\, \Kdoto_A \bigr), \\ 
g_{AB} &= r^2 \Omega_{AB} 
- \frac{1}{3} r^4 \bigl( \eq_7\, \Eq_{AB} - \bq_7\, \Bq_{AB} \bigr) 
+ \frac{5}{18} r^5 \bigl( \eq_8\, \Edotq_{AB} 
  - \bq_8\, \Bdotq_{AB} \bigr)  
- r^4\, \chi \partial_\phi \bigl( \ehatq_7\, \Eq_{AB} 
  - \bhatq_7\, \Bq_{AB} \bigr)
\nonumber \\ & \quad \mbox{} 
+ r^5\, \chi \partial_\phi \bigl( \ehatq_8\, \Edotq_{AB} 
  + \bhatq_8\, \Bdotq_{AB} \bigr)
- r^4 \bigl( \fo_7\, \Fo_{AB} - \ko_7\, \Ko_{AB} \bigr)
+ r^5 \bigl( \fo_8\, \Fdoto_{AB} + \ko_8\, \Kdoto_{AB} \bigr),
\end{align} 
\end{subequations} 
\ew
in which $\ehatq_n$, $\bhatq_n$, $\kd_n$, $\ko_n$, $\fd_n$, and
$\fo_n$ are functions of $r$ that are determined by solving the vacuum 
Einstein field equations. They are listed in Table~\ref{tab:radial} of this paper 
and in Table III of~\cite{Poisson:2014gka}.  

\begin{table*}
\begin{centering}
\caption{Radial functions appearing in the metric of
  Eq.~(\ref{metric}) , expressed in terms of $x = r/(2M)$.} 
\begin{ruledtabular} 
\begin{tabular}{l} 
$ \eq_2 = \frac{3(x-1)^2}{x^3}\, \ln(x) 
+ \frac{(x-1)(4x^4+5x^3-27x^2+7x+3)}{4x^5} $ \\ 
$ \eq_5 = \frac{2(x-1)}{x^2}\, \ln(x) 
+ \frac{(x-1)(6x^4+13x^3-15x^2-9x-3)}{6x^5} $ \\  
$ \eq_8 = \frac{3(2x^2-1)}{5x^3}\, \ln(x) 
+ \frac{(x-1)(5x^3+13x^2+4x-3)}{5x^4} $ \\ 
$ \bq_5 = \frac{2(x-1)}{x^2}\, \ln(x) 
+ \frac{(x-1)^2(6x^3+13x^2+4x+1)}{6x^5} $ \\ 
$ \bq_8 = \frac{3(2x^2-1)}{5x^3}\, \ln(x) 
+ \frac{(x-1)(5x^3+10x^2+x-1)}{5x^4} $ \\ 
\\
$ \ehatq_2 = \frac{(x-1)^2}{x^3} \mbox{dilog}(x) 
+ \frac{(x-1)^2}{2 x^3} \ln(x)^2 
- \frac{(x-1)(12x^2-9x+1)}{12 x^5} \ln(x) 
- \gam{q} \frac{4x+3}{48 x^4} 
+ \gamdot{q} \frac{1}{32 x^5} 
+ \frac{257}{108 x} - \frac{151}{27 x^2} + \frac{1643}{432 x^3} 
- \frac{31}{24 x^4} + \frac{97}{432 x^5} - \frac{1}{48 x^7} $ \\ 
$ \ehatq_5 = \frac{2(x-1)}{3 x^2} \mbox{dilog}(x)
+ \frac{x-1}{3 x^2} \ln(x)^2 
- \frac{12x^2-10x-1}{18 x^4} \ln(x) 
- \gamdot{q} \frac{2x+1}{48 x^5} 
+ \frac{257}{162 x} - \frac{335}{162 x^2} + \frac{20}{27 x^3} 
+ \frac{47}{324 x^4} + \frac{41}{648 x^5} - \frac{1}{18 x^6} $ \\ 
$ \ehatq_8 = \frac{2x^2-1}{6 x^3} \mbox{dilog}(x) 
+ \frac{2x^2-1}{12 x^3} \ln(x)^2 
- \frac{(4x+1)(3x-1)}{36 x^4} \ln(x) 
+ \gam{q} \frac{1}{72 x^3} 
- \gamdot{q} \frac{1}{48 x^4} 
+ \frac{257}{324x} - \frac{1}{8x^2} - \frac{37}{324x^3} 
+ \frac{101}{648x^4} + \frac{1}{72x^6} $ \\ 
$ \bhatq_5 = -\frac{2(x-1)}{3 x^2} \mbox{dilog}(x) 
- \frac{x-1}{3x^2} \ln(x)^2 
+ \frac{12x^2-10x-1}{18 x^4} \ln(x) 
+ \cc{q} \frac{1}{4 x^2} 
+ \frac{341}{216x} - \frac{31}{24x^2} - \frac{17}{27x^3}
+ \frac{8}{27x^4} + \frac{1}{9x^5} + \frac{1}{54x^6} $ \\ 
$ \bhatq_8 = -\frac{2x^2-1}{6 x^3} \mbox{dilog}(x) 
- \frac{2x^2-1}{12x^3} \ln(x)^2 
+ \frac{(4x+1)(3x-1)}{36 x^4} \ln(x) 
+ \ccdot{q} \frac{1}{8 x^3} 
+ \frac{341}{432x} +\frac{1}{8x^2} - \frac{13}{36x^3}
+ \frac{7}{108x^4} - \frac{1}{216x^6} $ \\ 
\\ 
$ \kd_2 = -\frac{(5x-1)(2x-1)(x-1)}{10x^5}\, \ln(x) 
- \cc{d} \frac{6x-1}{32x^5} 
+ \ccdot{d} \frac{1}{32 x^5} 
- \frac{11}{20x^2} + \frac{251}{120x^3} - \frac{34}{15x^4} 
+ \frac{1}{180x^5} - \frac{1}{60x^6} - \frac{1}{120x^7} $ \\ 
$ \kd_5 = - \frac{5x-4}{10x^3} \ln(x) 
- \cc{d} \frac{1}{32x^5} 
- \ccdot{d} \frac{1}{32x^5} 
- \frac{13}{24x^2} + \frac{14}{15x^3} - \frac{3}{20x^4}
+ \frac{19}{90x^5} - \frac{1}{120x^6} $ \\ 
$ \fd_5 = -\frac{5x-4}{10x^3}\, \ln(x) 
+ \gamdot{d} \frac{1}{4x^2} 
+ \frac{17}{15x^3} - \frac{3}{20x^4} - \frac{1}{40x^6} $ \\ 
\\ 
$ \ko_2 = \frac{(3x+1)(x-1)}{6x^5}\, \ln(x) 
+ \cc{o} \frac{1520x^5-3800x^4+3040x^3-660x^2+30x+3}{960x^5} 
+ \ccdot{o} \frac{1}{32x^5}
- \frac{1097}{36}   
+ \frac{5485}{72x} - \frac{544}{9x^2} + \frac{553}{36x^3} 
- \frac{19}{36x^4} - \frac{1}{360x^5} + \frac{1}{6x^6} 
+ \frac{1}{72x^7} $ \\ 
$ \ko_5 = -\frac{4x-5}{12x^4}\, \ln(x) 
+ \cc{o} \frac{2280x^5-3800x^4+1520x^3-15x-3}{1920x^5} 
- \ccdot{o} \frac{5x+1}{64x^5} 
- \frac{1097}{48} 
+ \frac{5485}{144x} - \frac{551}{36x^2} + \frac{1}{18x^3} 
+ \frac{79}{144x^4} - \frac{43}{240x^5} - \frac{7}{72x^6} $ \\ 
$ \ko_8 = -\frac{2x-1}{6x^4}\, \ln(x) 
+ \cc{o} \frac{760x^4-760x^3 +66x-3}{960x^4} 
- \ccdot{o} \frac{1}{32x^4} 
- \frac{1097}{72} 
+ \frac{1097}{72x} - \frac{1}{4x^2} - \frac{1237}{720x^3} 
+ \frac{151}{360x^4} + \frac{1}{36x^6} $ \\ 
$ \fo_5 = -\frac{4x-5}{12x^4}\, \ln(x) 
+ \gam{o} \frac{3x-5}{16x} 
- \frac{7949}{144} 
+ \frac{39745}{432x} - \frac{1000}{27x^2} - \frac{1}{9x^3} 
+ \frac{8}{9x^4} - \frac{1}{4x^5} - \frac{7}{24x^6} $ \\ 
$ \fo_8 = -\frac{2x-1}{6x^4}\, \ln(x) 
+ \gam{o} \frac{x-1}{8x} 
+ \gamdot{o} \frac{1}{8x^3} 
-\frac{7949}{216} 
+ \frac{7949}{216x} - \frac{1}{4x^2} + \frac{5}{9x^4} 
+ \frac{1}{12x^6} $  
\end{tabular}
\end{ruledtabular} 
\label{tab:radial} 
\end{centering}
\end{table*} 

As documented in~\cite{Poisson:2014gka}, the general solution for each radial
function involves two types of integration constants. The first
corresponds to a redefinition of a tidal multipole moment, and these 
constants can be set equal to zero without loss of generality. The
second type corresponds to the residual freedom of the light-cone
gauge, and these constants can be assigned arbitrarily without
altering the geometrical meaning of the coordinates. In~\cite{Poisson:2014gka}
the six gauge constants $\gam{d}$, $\gam{q}$, $\gam{o}$, $\cc{d}$,
$\cc{q}$, and $\cc{o}$ were eventually determined by anchoring the
coordinates to the null generators of the event horizon. We forego
this exercise here, and keep the constants arbitrary. In addition to
these, the new terms involving $\Edot_{ab}$ and $\Bdot_{ab}$ feature a
set of six additional constants denoted $\gamdot{d}$, $\gamdot{q}$,
$\gamdot{o}$, $\ccdot{d}$, $\ccdot{q}$, and $\ccdot{o}$; we trust that
this notation will not induce confusion, but state nevertheless that,
for example, $\ccdot{q}$ is {\it not} the time derivative of the
constant $\cc{q}$.  

We also note that the radial functions associated with the new
terms involving $\Edot_{ab}$ and $\Bdot_{ab}$ feature the dilogarithm 
function, defined by 
\begin{equation} 
\mbox{dilog}(x) = -\int_1^x \frac{\ln t}{t-1}\, dt. 
\end{equation}
%

\subsection{Teukolsky function} 
\label{sec:teukolsky} 

We can now use the metric obtained in Sec.~\ref{sec:metric} to
calculate the NP scalar 
\begin{equation} 
\psi_0 = -C_{\alpha\gamma\beta\delta} 
k^\alpha m^\gamma k^\beta m^\delta 
\label{psi0_def} 
\end{equation} 
of a slowly rotating, tidally deformed BH. Here
$C_{\alpha\gamma\beta\delta}$ is the NP tensor of the perturbed
spacetime, and $k^\alpha$ and $m^\alpha$ are two members of a null
tetrad required to be aligned with the Kinnersley tetrad in the background
spacetime. By virtue of the algebraic structure of the NP tensor in
the background spacetime, the computation of $\psi_0$ requires only
the {\it perturbation} of the NP tensor, and the {\it background}
tetrad vectors, which are given by
\begin{align} 
k^\alpha &= \biggl[ \frac{2}{f_s}, 1, 0, \frac{\chi M (1+2M/r)}{r^2 f_s}
\biggr], \qquad \\
m^\alpha &= \frac{1}{\sqrt{2}\, r} \biggl( 1 - i \frac{\chi M}{r}
\cos\theta \biggr) \biggl[ i \chi M\sin\theta, 0, 1,
\frac{i}{\sin\theta} \biggr] 
\label{tetrad} 
\end{align} 
in $(v,r,\theta,\phi)$ coordinates. These expressions are valid to
first order in $\chi$. 

Using the  spin-weighted spherical harmonics 
$\mbox{}_2 Y_{\ell}^m(\theta,\psi)$, with the explicit form employed in~\cite{Chatziioannou:2012gq} and noting that the azimuthal dependence is described by $\psi$, as defined by Eq.~(\ref{phi_vs_psi}), we find that the NP scalar can be decomposed as 
\begin{subequations} 
\label{psi0_1} 
\begin{align} 
\psi_0(v,r,\theta,\phi) &= \sum_{m=-2}^{2} \psi_0^{m}(v,r,\theta,\phi),
\\ 
\psi_0^{m}(v,r,\theta,\phi) &= 
{\cal R}_{2}^m(v,r)\, \mbox{}_2 Y_{2}^m(\theta,\psi), 
\end{align} 
\end{subequations} 
with 

\begin{align} 
{\cal R}_2^m(v,r) &= \alpha_m(v)\, {\cal P}_2^m(r)
+ \dot{\alpha}_m(v)\, M {\cal Q}_2^m(r) \nn 
\\
&+ i \beta_m(v)\, {\cal S}_2^m(r)  
+ i \dot{\beta}_m(v)\, M {\cal T}_2^m(r), \label{psi0_2}, 
\end{align} 
where $\alpha_m$ and $\beta_m$ are defined in terms of $\Eq_\m$ and $\Bq_\m$ in Eqs.~(28) of \cite{Chatziioannou:2012gq}, and where the radial functions are  given by 
\begin{subequations} 
\label{psi0_3} 
\begin{align} 
{\cal P}_2^m &= -1 
- \frac{(2y-1)(6y^2-6y-1)}{12(y-1)^2y^2} i m \chi, 
\\ 
{\cal Q}_2^m &= 2 \ln(y)
+ \frac{4y^5-2y^4-26y^3 +31y^2-4y-1}{6(y-1)^2y^2} 
\nonumber 
\\ &  \mbox{} 
+ \biggl[ 2 \mbox{dilog}(y) + \ln(y)^2 
\\
&+ \frac{598y^4-1214y^3+361y^2+204y+33}{108(y-1)^2y^2} \biggr] 
i m \chi, 
\\  
{\cal S}_2^m &= -1 
- \frac{(2y-1)(6y^2-6y-1)}{12(y-1)^2y^2} i m \chi, 
\\ 
{\cal T}_2^m &= 
2 \ln(y)
+ \frac{4y^5-2y^4-26y^3 +31y^2-4y-1}{6(y-1)^2y^2} 
\nonumber 
\\
 &  \mbox{} 
+ \biggl[ 2 \mbox{dilog}(y) + \ln(y)^2 
\\
&- \frac{285y^4-558y^3+443y^2-136y-22}{72(y-1)^2y^2} \biggr] 
i m \chi, 
\end{align} 
\end{subequations} 
where $y \equiv r/(2M)$. The decomposition of $\psi_0^m$ includes terms with $\ell = 2$ and $\ell = 3$, but the latter were not displayed here because they do not contribute to the horizon fluxes. We have verified that $\psi_0$ (with all terms included) satisfies the Teukolsky equation linearized with respect to $\chi$. 

The expressions displayed in Eq.~\eqref{psi0_3} imply that the
asymptotic behavior of the radial functions is given by 
\begin{subequations}
\label{radial_asymp}
\begin{align}
{\cal{P}}_2^m &= {\cal{S}}_2^m = -1 -2 i m \chi \frac{M}{r} + {\cal{O}}(M^2/r^2), \\
{\cal{Q}}_2^m & = \frac{r}{3M} + 2 \ln{\frac{r}{2M}} + 1 - \left(\frac{\pi^2}{3}-\frac{299}{54}\right) i m \chi + {\cal{O}}(M/r),\\
{\cal{T}}_2^m & = \frac{r}{3M} + 2 \ln{\frac{r}{2M}} + 1 - \left(\frac{\pi^2}{3}+\frac{95}{24}\right) i m \chi + {\cal{O}}(M/r).
\end{align}
\end{subequations}
The constant terms in these expressions, including the terms proportional to $im\chi$, are important for our purposes, because they determine the overall normalization of the Teukolsky function. We wish to call attention to the $\pi^2$ terms, and recall the observation made in Sec.~\ref{emri}, that our final expressions for the fluxes disagree with those obtained in~\cite{Tagoshi:1997jy} for the test-particle limit. The discrepancy, given in Eq.~\eqref{error}, contains a term proportional to $\pi^2$, while no such term is present in the test-particle result. The asymptotic behavior derived in Eq.~\eqref{radial_asymp} is the first introduction of factors of $\pi^2$ in our calculation\footnote{The factor of $\pi^2$ arises from the asymptotic behavior of the dilog function in Eq.~\eqref{psi0_3}.}, and this indeed happens for all $m \neq 0$ modes. This leads us to suspect that the discrepancy might originate in the asymptotic behavior of the Teukolsky function; see Sec.~\ref{matchingcheck} though for a detailed defense of the above calculation.

\section{Mano-Suzuki-Takasugi radial function} 
\label{sec:MST} 

In order to test the robustness of our solution to the Teukolsky equation, we calculate the energy flux using the series solution obtained in~\cite{Mano:1996vt}  (hereafter referred to as MST) rather than Eqs.~\eqref{0-sol} and~\eqref{1st-sol}. We then use the results of Sec.~\ref{sec:matching} and Appendix~\ref{sec:slowrot} to normalize the MST radial function. The resulting energy flux is unaltered from Eq.~\eqref{Mdott}; it suffers from the same discrepancy from the results of~\cite{Tagoshi:1997jy} indicating that our solution to the Teukolsky equation is robust.

The (exact) Teukolsky equation is written in Kerr coordinates $(v,r,\theta,\psi)$, and each mode of the NP scalar is decomposed as 
\begin{equation} 
(\tilde{\psi}_0)_\ell^m = R^m_\ell(r) S^m_\ell(\theta) e^{i m \psi},  
\label{psi0_vs_RS} 
\end{equation}
with a tilde indicating a frequency-domain function. The complete
function is obtained by multiplying by $e^{-i\omega v}$, and summing
over $\ell$ and $m$. To integrate the Teukolsky equation we follow
MST and define
\begin{equation} 
\kappa  = \sqrt{1-\chi^2}, \qquad 
\epsilon = 2 M \omega, \qquad 
\tau = (\epsilon-m\chi)/\kappa, 
\end{equation} 
replace $r$ with a new independent variable $\xi$ defined by 
\begin{equation} 
r = M(1 + \kappa - 2\kappa \xi),  
\label{x_def} 
\end{equation}
and replace $R^m_\ell(r)$ with a new dependent variable $p^m_{\ell}(\xi)$ 
defined by
\begin{equation} 
R^m_{\ell} = N^m_{\ell} (-\xi)^{-s} (1-\xi)^{i(\epsilon-\tau)} p^m_{\ell}(\xi),  
\label{R_vs_p} 
\end{equation} 
where $N^m_{\ell}$ is a normalization constant and $s=+2$. It should be noted
that the range $r \geq r_+ = M(1+\kappa)$ corresponds to $\xi \leq 0$.  

The function $p^m_{\ell}(\xi)$ is expressed in MST as a sum of
hypergeometric functions, 
\begin{equation}
p^m_{\ell}= \!\!\!\sum_{n=-\infty}^{\infty} \!\!\!A_n(\nu) 
F(n+\nu+1-i\tau, \!-n-\nu-i\tau; \!1-s-i\epsilon-i\tau; \xi),
\label{p_def} 
\end{equation} 
where the coefficients $A_n(\nu)$ satisfy a three-point recurrence 
relation ($A_0$ can be set equal to unity without loss of generality),
and $\nu$ is a generalized angular-momentum parameter defined to
ensure that the sum converges. An alternative representation of $p(\xi)$
is   
\begin{equation} 
p^m_{\ell}(\xi) = q^m_{\ell}(\nu ; \xi) + q^m_{\ell}(-\nu-1; \xi) 
\label{p_vs_q} 
\end{equation} 
with 
\begin{align} 
&q^m_{\ell}(\nu ; \xi) = \!\!\!\!\!\sum_{n=-\infty}^{\infty} \!\!\!A_n(\nu) 
\frac{\Gamma(1-s-i\epsilon-i\tau) \Gamma(2n+2\nu+1)} 
{\Gamma(n\!+\nu\!+1\!-i\tau) \Gamma(n\!+\nu\!+1\!-s\!-i\epsilon)} \nonumber 
\\ 
&\! \times \!
(-\xi)^{n+\nu+i\tau} 
F(\!-n\!-\nu\!-i\tau, \!-n\!-\nu\!+\!s\!+\!i\epsilon; \!-2n\!-\!2\nu; \!1/\xi). 
\label{q_def} 
\end{align} 
Equation (\ref{p_def}) is useful when one is interested in the
behavior of the radial function near $\xi=0$ ($r = r_+$). The
alternative form of Eqs.~(\ref{p_vs_q}) and (\ref{q_def}) is useful
when $-\xi \gg 1$ ($r/M \gg 1$). 

For our purposes it is sufficient to set $\ell = 2$ and expand
$R^m_{2}(r)$ to first order in $\epsilon$. We have $\nu = 2
- \frac{107}{210} \epsilon^2 +  {\cal{O}}(\epsilon^3)$, 
\begin{align} 
A_{-3} &= \frac{28}{107} m \chi (\kappa - i m \chi)
  (2\kappa - i m \chi) \epsilon + {\cal{O}}(\epsilon^2), \\ 
A_{-2} &= -\frac{28}{107} m \chi (\kappa - i m \chi)
  (2\kappa - i m \chi) \epsilon +{\cal{O}}(\epsilon^2), \\ 
A_{-1} &= \frac{2i}{5} (2\kappa - i m \chi) \epsilon 
+ {\cal{O}}(\epsilon^2), \\ 
A_0 &= 1, \\
A_1 &= \frac{i}{90} (3\kappa+i m \chi) \epsilon 
+ {\cal{O}}(\epsilon^2),    
\end{align} 
and all other coefficients are higher order in $\epsilon$. These
results can be inserted in Eq.~(\ref{p_def}) to obtain $p^m_{2}(\xi)$ to first order in $\epsilon$. The
angular functions are known also to admit an expansion in 
$\epsilon$, given schematically by 
\begin{align} 
S_2^m(\theta) e^{i m \psi} 
&\!=\! \mbox{}_2 Y_2^m(\theta,\psi)  
\!+\! \epsilon \Bigl[\! \mbox{}_2 Y_{3}^m(\theta,\psi)  
 \mu_+ \!+\! \mbox{}_2 Y_{1}^m(\theta,\psi) \mu_-\! \Bigr] \nn
\\
&+ {\cal{O}}(\epsilon^2),  
\end{align} 
where $\mu_\pm$ are numbers proportional to $\chi$~\cite{Chatziioannou:2012gq}. Making the
substitution in Eq.~(\ref{psi0_vs_RS}) gives 
\begin{align} 
(\tilde{\psi}_0)_2^m
&= {\scr R}^m_2(r)\, \mbox{}_2 Y_2^m(\theta,\psi) 
+  {\scr R}^m_{3}(r)\, \mbox{}_2 Y_{3}^m(\theta,\psi)\nn
\\
&+  {\scr R}^m_{1}(r)\, \mbox{}_2 Y_{1}^m(\theta,\psi)
+ {\cal{O}}(\epsilon^2), 
\end{align} 
where ${\scr R}^m_2(r) = {\cal{O}}(1) + {\cal{O}}(\epsilon)$ is equal to the radial 
function $R^m_{2}$ expanded to first order in $\epsilon$, while 
${\scr R}^m_{2\pm 1}(r) = {\cal{O}}(\epsilon)$ are constructed from $R^m_{2}$
(truncated to order $\epsilon^0$) and $\mu_\pm$.    

To normalize the radial function we examine the regime $r/M
\gg 1$. Making the substitutions in Eqs.~(\ref{R_vs_p}),
(\ref{p_vs_q}), and (\ref{q_def}), and making use of Eq.~(\ref{x_def}), we find  
\begin{equation} 
(\tilde{\psi}_0)_2^m
\sim {\scr R}_2^m(r)\, \mbox{}_2 Y_2^m(\theta,\psi) 
+  {\scr R}_3^m(r)\, \mbox{}_2 Y_3^m(\theta,\psi)
+ {\cal{O}}(M^2 \omega^2),  
\label{psi0_expanded1} 
\end{equation} 
with 
\begin{align} 
{\scr R}_2^m(r) &= -Z_2^m \Biggl\{ 1 + \frac{i}{3} \omega r \biggl[ 1   
+ \frac{M}{r} \biggl(  6 \ln \frac{r}{2M} - 1 \nn
\\
&+ \frac{5}{3} i m \chi \biggr) \biggr] + {\cal{O}}(M/r, M^2 \omega^2) \Biggr\} 
\label{psi0_expanded2} 
\end{align} 
where $Z^m_2$ is a new normalization related to $N^m_2$ by 
\begin{equation} 
Z ^m_2= -N^m_2 \frac{24\, \Gamma(-1-i\epsilon-i\tau)} 
{\Gamma(3-i\tau) \Gamma(1-i\epsilon)} \kappa^{-i\epsilon},
\label{Z_vs_N} 
\end{equation} 
and ${\scr R}_3^m \propto i M\omega Z_2^m$. Notice that the asymptotic
behavior of the radial function is linear in $\chi$, enabling us to use the asymptotic value of the radial function derived through a first-order-in-$\chi$ metric of Appendix~\ref{sec:slowrot}.

Equations (\ref{psi0_expanded1}) and (\ref{psi0_expanded2}) can
now be compared with Eqs.~(\ref{psi0_1}), (\ref{psi0_2}), and
(\ref{radial_asymp}) to determine the amplitude $Z_2^m$ in relation to
$\tilde{\alpha}_m(\omega)$ and $\tilde{\beta}_m(\omega)$, the Fourier
transforms of the time-domain tidal moments $\alpha_m(v)$ and
$\beta_m(v)$, respectively. While the radial functions 
${\scr R}_2^m(r)$ and ${\cal R}_2^m(r)$ are formally distinct ---  the
first is valid to all orders in $\chi$, while the second is linearized
with respect to $\chi$ --- they can nevertheless be identified in the
asymptotic regime, which is insensitive to higher-order terms in
$\chi$. Simple algebra then yields 
\begin{equation} 
Z_2^m = ( 1 - i \Gamma_1 M \omega ) \tilde{\alpha}_m 
+ ( 1 - i \Gamma_2 M \omega ) i \tilde{\beta}_m   
+ {\cal{O}}(M^2 \omega^2), 
\label{Z_vs_alphabeta1} 
\end{equation} 
with 
\begin{equation} 
\Gamma_1\! =\! -\frac{4}{3} + \biggl( \frac{\pi^2}{3} 
- \frac{269}{54} \biggr) i m \chi, \quad 
\Gamma_2 \!=\! -\frac{4}{3} + \biggl( \frac{\pi^2}{3} 
+ \frac{325}{72} \biggr) i m \chi. 
\label{Z_vs_alphabeta2} 
\end{equation} 
With $N_2^m$ related to $Z_2^m$ through Eq.~(\ref{Z_vs_N}), the
normalization of the MST radial function is now determined. Using this form for the Teukolsky function and following the same steps as Sec.~\ref{hf}, we again arrive at Eq.~\eqref{Mdott}. Both methods to solve the Teukolsky equation produce the same discrepancy with the results as~\cite{Tagoshi:1997jy}.  

\bibliography{master.bib}

\end{document}